# Diffraction-Free Bloch Surface Waves

Ruxue Wang[1‡], Yong Wang[1‡], Douguo Zhang[1]*, Guangyuan Si[2], Liangfu Zhu[1], Luping Du[3], Shanshan Kou[4], Ramachandram Badugu[5], Mary Rosenfeld[5], Jiao Lin[2,3], Pei Wang[1], Hai Ming[1], Xiaocong Yuan[3], and Joseph R. Lakowicz[5]

[1] Department of Optics and Optical Engineering, University of Science and Technology of China, Hefei, Anhui, 230026, China

[2] School of Engineering, RMIT University, Melbourne, VIC 3001, Australia

[3] Nanophotonics Research Centre, Shenzhen University & Key Laboratory of Optoelectronic Devices and Systems of Ministry of Education and Guangdong Province, College of Optoelectronic Engineering, Shenzhen University, Shenzhen 518060, China.

[4] Department of Chemistry and Physics, La Trobe Institute for Molecular Sciences (LIMS), La Trobe University, Melbourne, Victoria 3086, Australia.

[5] Center for Fluorescence Spectroscopy, Department of Biochemistry and Molecular Biology, University of Maryland School of Medicine, Baltimore, MD 21201, United States

[‡]These authors contributed equally to this work.

*Correspondence and requests for materials should be addressed to D.G. Zhang. (Email: dgzhang@ustc.edu.cn)

**Abstract:**

In this letter, we demonstrate a novel diffraction-free Bloch surface wave (DF-BSW) sustained on all-dielectric multilayers that does not diffract after being passed through three obstacles or across a single mode fiber. It can propagate in a straight line for distances longer than 110 µm at a wavelength of 633 nm and could be applied as an in-plane optical virtual probe, both in air and in an aqueous environment. The ability to be used in water, its long diffraction-free distance, and its tolerance to multiple obstacles make this DF-BSW ideal for certain applications in areas such as the biological sciences, where many measurements are made on glass surfaces or for which an aqueous environment is required, and for high-speed interconnections between chips, where low loss is necessary. Specifically, the DF-BSW on the dielectric multilayer can be used to develop novel flow cytometry that is based on the surface wave, but not the free space beam, to detect the surface-bound targets.

Bloch surface waves (BSWs) are electromagnetic surface waves excited at the interface between a truncated periodic dielectric multilayer with a photonic band gap (PBG) and its surrounding medium. They can be considered to be the dielectric analogue of surface plasmon polaritons (SPPs), except with lower losses and longer propagation lengths [1-2]. Similar to SPPs, BSWs have been applied in nanoscale optical circuits, bio-sensing, gas sensing, fluorescence emission enhancement or sorting, and surface enhanced Raman scattering [3-7]. BSWs possess specific properties that differentiate them from SPPs. They are not subject to losses caused by metal absorption, which allows for BSW resonances with high quality factors and long propagation lengths [8]. There are many choices for the dielectric materials for BSWs, which allows this dielectric multilayer to be used from deep ultraviolet (UV) to near-infrared (NIR) wavelengths [9]; meanwhile SPPs suffer considerably higher propagation losses in the UV and visible region. So, various methods have been proposed to increase the propagation lengths of SPPs [10-11], that is still not long enough for practical applications. Additionally, fluorophores can be quenched near metallic surfaces, which does not occur with dielectric surfaces. Further, the moderate localization of a BSW mode compared with that of a SPP mode is favorable for applications where a large volume of the adjacent material needs to interact with the dielectric multilayer.

Both SPPs and BSWs undergo diffraction in the plane of the interface, which will induce coupling losses between the on-chip components as the wave packet spreads laterally during propagation. In recent years, with the rapid development of plasmonic technology, many methods have been proposed to construct diffraction-free SPPs [12], such as plasmonic Airy beams (PABs) generated by spatial light modulators [13], in-plane diffraction from metallic hole arrays [14], Cosine-Gauss plasmon beams created by two sets of metallic gratings [15], and efficient manipulation of PABs in linear optical potentials produced by wedged metal–dielectric–metal structures [16]. The evolution from free-space three dimensional (3D) diffraction-free optical waves (such as Bessel and Airy beams) into two dimensional (2D) surface waves, is not only of fundamental interest, but could also facilitate the development of devices that utilize these waves. One of the main advantages of this technique is that 2D elements can have arbitrary shapes—something that is difficult to achieve in 3D [17]. This is also one of the reasons why research is shifting from metamaterials to metasurfaces [18-19]. Previously, little to no effort had been devoted to constructing diffraction-free surface waves in an aqueous environment, which is more favorable for biological applications. Many biological and clinical assays are performed on glass surfaces, including assays involving DNA [20], proteins [21], and HIV [22], as well as a wide range of immunoassays [23]. More recently these assays are being modified to use multiplexing based on multiple excitation and emission channels [24-25]. The ability to generate and control diffraction-free light waves on glass surfaces could have a large impact on the biosciences and result in new formats for diagnostic devices. Such applications are more difficult to achieve with metallic structures because of the high propagation losses at the metal/water interface. The propagation loss of a SPP at a metal/water interface is higher than that of a SPP at a metal/air interface, notably in the visible spectrum, such as for a wavelength of 633 nm. For a BSW with dielectric multilayers, the propagation losses are much lower (both in water and air), and their effective indices can be tuned by varying the thickness of the top dielectric layer [26]. Using an axicon (or conical lens element [27-29]) free-space Bessel beams, which are also a diffraction-free beam, can be generated; this has been reported in free space optics (3D optics). However, to the best of our knowledge, there are seldom reports describing how to realize diffraction-free BSW on a 2D optical surface, especially in aqueous environment and at visible light band. In the present report, we demonstrate both experimentally and theoretically that diffraction-free BSWs can be generated on an all-dielectric multilayer with simple dielectric gratings; we demonstrated this both in an air and aqueous environment, at a wavelength of 633 nm. Owing to the low loss of BSWs, diffraction-free behavior can be preserved even after the wave encounters either three obstacles or a single mode silicon fiber along the propagation path; this had not been demonstrated previously with SPPs. The ability of a DF-BSW to reform after an obstruction will facilitate their use in devices that require measurements at multiple locations.

The dielectric multilayer is made of alternating layers of $SiO_2$ and $Si_3N_4$, with the number of layers shown in Figure 1a. Except for the top $SiO_2$ layer, which was 450 nm thick (refractive index n = 1.46), the $SiO_2$ layers were 110 nm thick. The thickness of the $Si_3N_4$ (Si Rich+) layers was approximately 66 nm (refractive index n = 2.6). With this multilayer structure, BSWs can be populated at the water/$SiO_2$ interface (Supplementary Figure 1). Two dielectric gratings (Figure 1b) with a period of 460 nm were inscribed on the top $SiO_2$ layer with a focused ion beam (FIB); the grating's period was matched with the physical wavelength of the BSWs at the $SiO_2$/water interface (Figure 1b, which shows that the crossing angle of the two gratings were designed to be either 10° or 170°). Subsequently, the multilayer was coated with a drop of water. With the aid of the grating for momentum matching, BSWs were excited with a normally incident, focused Gaussian beam from a laser with linear polarization parallel to the Y-axis at a wavelength of 633 nm. The laser beam was focused onto the center of the two gratings. The plane BSWs of equal amplitude launched by the pair of gratings are expected to interfere constructively to form DF-BSWs for a distance $X_{max} = \omega_0/\sin(\vartheta) = 115$ μm, where $\vartheta$ was designed to be 5° and the beam waist of the incident beam was $\omega_0$, which was set to 10 µm. The length of the grating (D) was 30 µm and thus larger than the beam waist (Fig. 1c). It should be noted that the calculation of the maximum non-diffraction distance was based on the assumption that BSWs propagate without loss (from point S to point T, where point S was set to be the zero point of the X–Y axis). The optical field intensity distribution of the BSWs was then measured with a home-built leakage radiation microscope (LRM, Fig. 1d) [30].

Figure 2a shows a front focal plane (FFP) image of the LRM; in it, a significant main lobe and several parallel side lobes appear and propagate toward the right side. A numerical simulation of the electric filed distribution on the top surface of the dielectric layer (SiO$_2$/water interface, Figure 1a) was performed with the finite-difference time-domain (FDTD) method (Figure 2c), where the layer thicknesses, grating parameters, and incident wavelength were the same as those used in the experiments (Figure 1). Consistent with the LRM image, a strong main lobe and several weak parallel side lobes are visible in the simulated image. Because of the scattering caused by the roughness around the gratings, the intensity is brighter in the areas of excitation in the experimental image (Figure 2a) than in the simulated one (Figure 2c). To assess the diffraction-free properties, we plotted the intensity profiles along the white dashed lines at three distances X ($X$ = 50, 70, and 80 µm, where the position of the starting point S was set to be $X$ = 0 µm (Figure 1c); for clarity, only one dashed line is shown at $X$ = 50 µm) for both the experimental (Figure 2b) and simulated (Figure 2d) images. The beam presents a narrow profile, and the full-width at half-maximum of the main lobe was about 2 µm at all three distances (or $X$-positions) in both the numerical and experimental results. Similar to the main lobe, the transverse profiles (or beam shapes) of the side lobes were also preserved for the surface waves propagating along the propagating direction. For the cosine-Gauss SPP wave [15], only the main lobe appeared and retained diffraction-free owing to the high loss of the SPPs. Another difference is that BSWs can be populated with either transverse electric (TE) or transverse magnetic (TM) polarized light if the correct layer thicknesses are chosen [31], which yields even more polarization choices. In this work, the DF-BSW was excited with an incident beam with a polarization along the Y-axis (Figure 2a and 2b) causing the BSW to be TE polarized (the DF-BSW here cannot be excited with TM polarized light as demonstrated by the image in Supplementary Figure 2). In contrast, for diffraction-free SPPs, the excitation field and propagation field must be TM polarized (SPPs can only be generated with TM polarized light, as shown in Supplementary Figure 3, where the incident polarization direction is along the *X*-axis). For comparison, the electric field distribution of the diffraction-free SPPs at the silver/water interface originating from the two gratings was also simulated with FDTD at the same wavelength, i.e. at 633 nm (Supplementary Figure 3); this shows that the propagation distance is considerably reduced.

To clearly show the formation of the DF-BSW, the laser beam was focused onto a single grating where plane BSWs that propagate both toward the right and left sides are generated (Figure 2e). The corresponding back focal plane (BFP) image (Figure 2f) shows three points. The center one represents the transmitted laser beam. Normally, the dielectric multilayer would reflect all light with a wavelength of 633 nm because of its PBG, meaning no light would pass through the multilayer at normal incidence. Because of the inscribed gratings on the top $SiO_2$ layer (the gratings can be seen as defects in the multilayer), the PBG of the multilayer is broken and therefore the light can be transmitted, which results in the center spot on the BFP image. The BFP image represents the distribution of the wave-vector or divergence of the corresponding optical waves. If the spot on the BFP image is large, it means that the corresponding wave propagates in a wide range of directions (i.e. it has a large divergence). Hence, the known numerical aperture (NA) of the objective (NA of

1.49) and the size of the center spot divergence can be derived at approximately 2°, meaning that the exciting light is essentially of normal incidence. Similarly, the left and right spots correspond to BSWs from a single grating; the BSWs propagate toward both the left and the right side. From the BFP image, we can deduce that the propagation directions are both perpendicular to the long axis of the grating. The two spots in the image are very small, corresponding to an angular divergence of approximately 3°; hence, they can be seen as belonging to an approximately planar surface wave (Figure 1c). The distance between the right and left spots indicates the wavenumber (or effective index) of the excited BSW. The larger the distance, the larger the wavenumber is. The wavenumber was derived to be 1.38 $K_0$ in this case, which is consistent with the findings from the white light BFP image (Supplementary Figure 1).

For comparison, the BFP image shows two pairs of spots (Figure 2g, relative to the FFP image in Figure 2a). Two dashed lines connecting a pair of spots both pass through the center spot. Based on the above description, the two pairs of spots (on each dashed line) demonstrate that the two gratings each generate a BSW. Owing to the different orientations of the gratings (their cross-angle was either 170° or 10°, Figure 1), the propagation direction (illustrated by the two dashed lines) of the BSWs originating from each grating will be different, which results in the four spots visible in Figure 2g. The cross-angle of the launched planar BSWs, represented by the cross-angle of the two dashed lines, was 10°, which is consistent with the cross-angle of the two designed gratings (Figure 1b, $2\vartheta$ = 10°; the propagation direction of the launched BSW was perpendicular to the grating). The BSWs launched by each grating interfere when they overlap spatially, as illustrated in Figure 1c (the BSWs propagating to the right side). Our BFP images confirm the formation of the diffraction-free surface waves, which originate from the interference of the two planar surface waves caused by the one-dimensional gratings. This information cannot be resolved only from the FFP image (Figure 2g).

The penetration of the DF-BSW into water was simulated as shown in Supplementary Figure 4 (which shows the electric-field distribution in the *Y-Z* plane at different *X* positions). The penetration depth in water can reach about 400 nm, which can be increased or decreased by either decreasing or increasing the top $SiO_2$ layer, respectively [26]. The penetration of the DF-BSW into water can be used for fluorescence imaging. The dye-labelled cells are suspended in cell culture media, which are aqueous solutions. To demonstrate this, we doped the water with the dye Fluorescent red 646 reactive (from Fluka), which can be excited at a wavelength of 633 nm and which fluorescence at 660 nm. The fluorescence image (Figure 2h) shows that the fluorophores along the main lobe of the DF-BSW can be excited. This phenomenon suggests that the DF-BSW can work as an in-plane optical virtual probe for fluorescence based detection or sensing. For example, if a labeled cell in a microfluidic system flows across the DF-BSW, it will be excited and can then be counted or detected or sorted. Here, the DF-BSW is trapped on the surface and propagates far away from the focal point, so there is little contribution from the directly incident light, resulting in a low background noise. What is more, given the small distance of the evanescent penetration compared with the free space laser beam, the DF-BSW has the potential to be used for single molecule detection where the molecules are located far from the incident light. The fluorescence coupling efficiency appears to be low based on the relative brightness of the directly excited fluorescence and the DF-BSW. This may be the result of having used a thick layer of the fluorophore containing water, which may reduce surface-selective fluorophore binding.

An important advantage of this low loss dielectric multilayer is that the DF-BSWs can be preserved even after passing several obstacles (such as the three obstacles shown in Figure 3a, which were 2 µm × 2 µm square holes positioned at $X$ = 16 µm, 34 µm, and 52 µm, respectively); this had not previously been demonstrated experimentally for any diffraction-free SPPs. We conclude that one reason this had not previously been shown is the high absorption losses in metal; another reason being the high scattering loss of SPP waves caused by obstacles. Figure 3b and 3c clearly illustrate that the main lobe of the BSWs still propagates along a straight line after passing the three obstacles, even when the distance between the starting point (S) and the third obstacle reaches around 52 µm. The intensity profiles at $X$ = 90 µm and 110 µm (Figure 3d and 3e) clearly demonstrate that the diffraction-free behavior of the main lobe is maintained after the three obstacles (it also was named as self-healing property). The diffraction-free distance thus approaches the maximum theoretical value ($Xmax = \omega_0/\sin(\vartheta)$ = 115 µm ), which demonstrates that a BSW can propagate much further than a SPP. The main lobe becomes more distinct than the side lobe after the obstacles—this is true in both in the simulated and experimental images. In contrast, the main lobe is weaker than the side lobe at long distances (such as at $X$ = 100 µm, Figure 2a and 2c) in the absence of obstacles. What is more, a comparison of Figures 3c and 2c shows that BSWs propagate further when obstacles are present. In Figure 2c the maximum intensity of the main lobe occurs around $X$ = 40 µm, whereas in Figure 3c, it occurs around $X$ = 90 µm. These properties potentially provide a way to obtain a strong main lobe for the diffraction-free surface waves. In addition to the regular obstacles designed along the travel path of the main lobe, a single mode silicon fiber with a diameter of about 9 µm and longer than 50 µm was placed on the multilayer (Figure 3f). While one might expect that this would block the propagation of both the BSW's main and side lobes, Figure 3g shows that the main and side lobes of the BSW preserve their diffraction-free property after passing through the fiber. To the best of our knowledge, there is no previous report of diffraction-free SPPs that can pass through an obstacle that is so large.

If we look at the self-healing ability of the DF-BSWs from a different point of view, another novel application can be proposed. Both the square hole and round fiber obstacles induce an optical surface field distortion along the main lobe of the DF-BSW (there are three dark spots at the obstacles' locations in Figure 3b, marked with dashed lines), which provide an approach to detect or count the particles passing across the BSW. When a serial of particles (or cells) flow across the non-diffraction BSW, they will induce a distortion of the surface beam, which can be used to count particles or cells. The principle is similar to that of the widely used flow cytometer (light scattering is widely used in flow cytometry to register when a cell is located in the path of a free-space beam); however, using our method we would be able to only count the particles or cells close to the surface because of the small penetration depth of the surface wave. This may be useful though, as it may reduce the background noise, particularly for some surface-bound applications. The self-healing ability and long propagation distance of the DF-BSWs can be used for parallel detection of particles, such as detecting three or even more obstacles simultaneously, as demonstrated in Figure 3b. Unlike the proposal to count fluorophore label cells, this method is based on the scattering of DF-BSWs and represents a label-free approach.

Similar to a BSW at the dielectric/water interface, a BSW at a dielectric/air interface can also be generated on a dielectric multilayer (Supplementary Figure 5a). In this case, a BSW with a smaller wavenumber than when using water is observed (smaller dark ring in the BFP images, Supplementary Figure 5b). With the aid of two gratings (Supplementary Figure 5c, grating period of 590 nm to match the wavelength of the BSW at the air/$SiO_2$ interface), DF-BSWs are generated as shown in the FFP image of the LRM (Figure 4a, captured with a 60× objective with a NA of 1.20), in which both the main lobe and side lobes appear simultaneously. The BFP image (Figure 4b) also shows two pairs of spots on the left and right sides, where the cross-angle is also 10°. When a square obstacle with a size of 2 μm ×2 μm was placed in the propagating path (Figure 4c), the BSWs showed self-healing, and the main lobe became more dominant after the obstacle (Figure 4d). The DF-BSWs in air can thus be used to detect atmospheric particles that flow close to the surface of the dielectric multilayer. The advantage of this approach is that the particle can cause a signal on a dark background, which is much easier to detect than a small decrease in a high intensity signal.

From Figure 4a it can be seen that the propagation of the DF-BSW can exceed a distance of 110 μm from starting point S at a wavelength of 633 nm. For a SPP with a wavelength of 785 nm, the diffraction-free surface waves can only travel 80 μm [15]; this distance will be further decreased if the incident wavelength were decreased to 633 nm, i.e. to the same wavelength as used for the BSW (Supplementary Figure 3). As described by the equation $X_{max} = \omega_0/\sin(\vartheta)$, we can increase the non-diffraction distance $X_{max}$ by increasing $\omega_0$ or by decreasing $\vartheta$, which does not change the intrinsic propagation loss of the SPPs or BSWs—however, the SPPs and BSWs may lose their energy before they reach the terminal point T, in which case this equation cannot be used. The diffraction-free distance of the surface waves generated by the intersecting gratings is strongly correlated with the intrinsic propagation loss of the plane BSWs or SPPs. For a SPP at an Ag/water interface, the theoretical propagation length is approximately 8.5 μm for a wavelength of 633 nm, regardless of the roughness of the silver film and the leakage radiation loss used for imaging [32]. Whereas, for the BSW shown on Figure 2e, the experimental propagation length can be calibrated to be as long as 38 μm at the same wavelength, based on intensity trace fitting with an exponential function. The smaller the loss of the BSWs compared with that of SPPs, the more favorable it is for increasing the real diffraction-free distance and also its tolerance to obstacles, as demonstrated in Figures 2, 3, and 4. It should be noted in the discussions of Figures 2, 3, and 4, we do not use the definition "propagation length" for the DF-BSW, because the intensity decay along the X-axis (propagation direction of the DF-BSW) is not exponential (its maximum intensity appears at nearly the center of the BD-BSW beam). Whereas, for a BSW excited by a single grating (Figure 2e), the excited BSW propagating to the left or right is a diffracting beam and is exponentially dampened during propagation.

In conclusion, we have experimentally and numerically demonstrated DF-BSWs for the first time, in water environment and at visible light band. The lower loss of the dielectric multilayer compared with a metal film, especially in the visible spectrum, makes BSWs more favorable than SPPs for the formation of diffraction-free surface waves, with the advantage of being diffraction-free over long distances and can be re-formed after passing through larger or smaller obstacles. Because of these unique advantages, DF-BSWs have potential applications in areas such as on-chip optical circuits [33-34], where the signal may encounter obstacles as the devices become more complex. DF-BSWs may also find use for the optical nano-manipulation of molecules in liquids [35-36] or as a light source for cell counting in flow cytometry [37]. Since a DF-BSW has an evanescent field, DF-BSWs can be used for cell imaging for cells located on surfaces as an alternative to total internal reflection, which requires a coupling prism [38]. Additionally, with the exception of the DF-BSWs demonstrated here, all other intensity or shape controllable surface waves, such as those generated by linearly focusing beams [39], de-multiplexing waves [40], in-plane Airy beams [14], trapping light that mimics gravitational lensing [41], and wave-front shaping by exploiting general relativity (GR) and its effects [42], can be attained with our platform, which consists of all-dielectric multilayers with water or air as the surrounding medium. We believe that our findings will inspire further intriguing phenomena to be discovered and that additional practical applications in beam engineering and nanophotonic manipulations will be uncovered.

**Methods:**

**Leakage radiation microscopy**

The incident laser beam (633 nm wavelength) was coupled into a single mode fiber and then collimated by an optical expander. The collimated beam was focused by a lens onto the dielectric gratings (Figure 1d, with a beam waist of 10 μm). An oil or water immersed objective (for oil: 100× magnification and NA of 1.49 from Nikon for a BSW at a dielectric/water interface, see Figures 2 and 3 as well as Supplementary Figure 2; for water: 60× magnification and NA of 1.20 From Nikon for a BSW at a dielectric/water interface, see Figure 4) was used to collect the leakage signals of the BSWs. By changing the position and focal length of the tube lens, either the FFP or BFP of the objective can be imaged by a CCD camera (Retiga 6000 from QImaging). A polarizer and a half wave plate can be inserted between the fiber-expander and the top focal lens to tune the incident polarization.

**Fabrication of the dielectric multilayers and gratings**

The dielectric multilayers were fabricated via plasma-enhanced chemical vapor deposition (PECVD, Oxford System 100) of $SiO_2$ and $Si_3N_4$ on a standard microscope cover glass (0.17 mm thickness) at a vacuum pressure of <0.1 mTorr and a temperature of 300 °C. For the BSWs in water, the refractive index of the (high index) dielectric layer $Si_3N_4$ (Si Rich+) was n = 2.60, and for the BSWs in air it was n = 2.14. The low refractive index dielectric layer was $SiO_2$ (n = 1.46). The thicknesses of these layers are shown in Figure 1a and in Supplementary Figure 5a, respectively. There were fourteen layers in total. The dielectric multilayer was then coated with a gold film of 20 nm thickness by magnetron sputtering (Sputter-Lesker Lab18); this gold film was then used to create two gratings on the top $SiO_2$ layer using FIB (Helios NanoLab 650) lithography. After the lithography step, the gold film was removed with aqua regia, and then the dielectric multilayer was cleaned with acetone and then with nanopure deionized water and dried with an $N_2$ stream.

**Acknowledgements**

This work was supported by MOST (2013CBA01703, 2016YFA0200601), NSFC (61427818, 11374286, 61622504), and the Science and Technological Fund of Anhui Province for Outstanding Youth (1608085J02). This work was also supported by grants from the National Institute of Health (GM107986, EB006521, EB018959, and OD019975). This work was partially carried out at the USTC Center for Micro and Nanoscale Research and Fabrication, and we thank Xiaolei Wen, Linjun Wang, and Yu Wei for their help on the micro/nano fabrication steps.

**Author contributions**

R.X.W and L.F.Z carried out the optical experiments, R.X.W, and G.Y.S fabricated the dielectric multilayer structures and gratings. Y.W. and R.X.W performed the numerical calculations. S.S.K., L.P.D, R.B., J.L., P.W., X.-C.Y., and J.R.L. assisted in analyzing the results and drafting the manuscript. D.G.Z. initiated the work and drafted the manuscript. All authors discussed the results and contributed to the writing and editing of the manuscript.

**Competing financial interests:** The authors declare no competing financial interests.

**Data availability**

The data that support the findings of this study are available from the corresponding authors upon reasonable request.

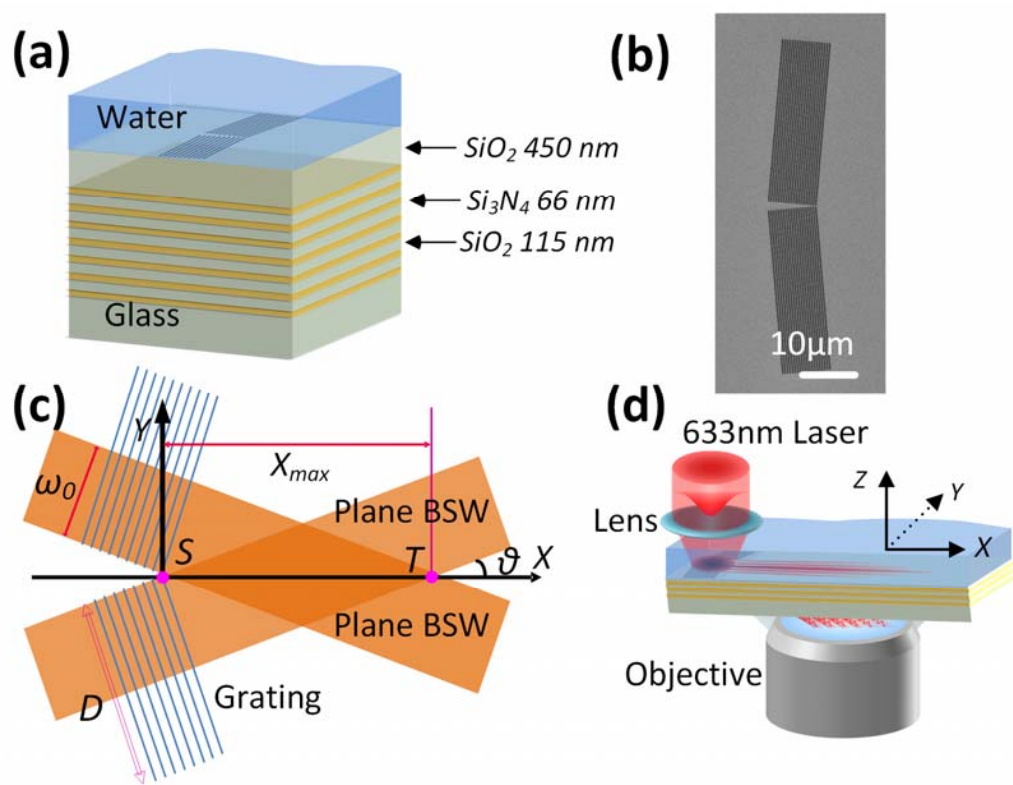

**Figure 1: | Schematic diagram of the samples and of the optical setup.** (a) Dielectric multilayer with intersecting gratings fabricated on the top $SiO_2$ layer. The top surface of the sample was immersed in deionized water. (b) A scanning electron microscope (SEM) image of the grating in which the cross-angle of the two gratings is 170°. The period of the gratings was 460 nm. (c) Formation of the DF-BSW. The cross-angle of the two gratings is defined as ($\pi-2\vartheta$ = 170°); $X_{max}$ represents the longest diffraction-free distance; D represents the length of the grating; $\omega_0$ is the beam waist; and S represents the zero point of the X–Y axis. (d) Optical setup of the LRM. The Gaussian beam was focused onto the two gratings and then launches the DF-BSW, whose leakage radiation is collected by the objective.

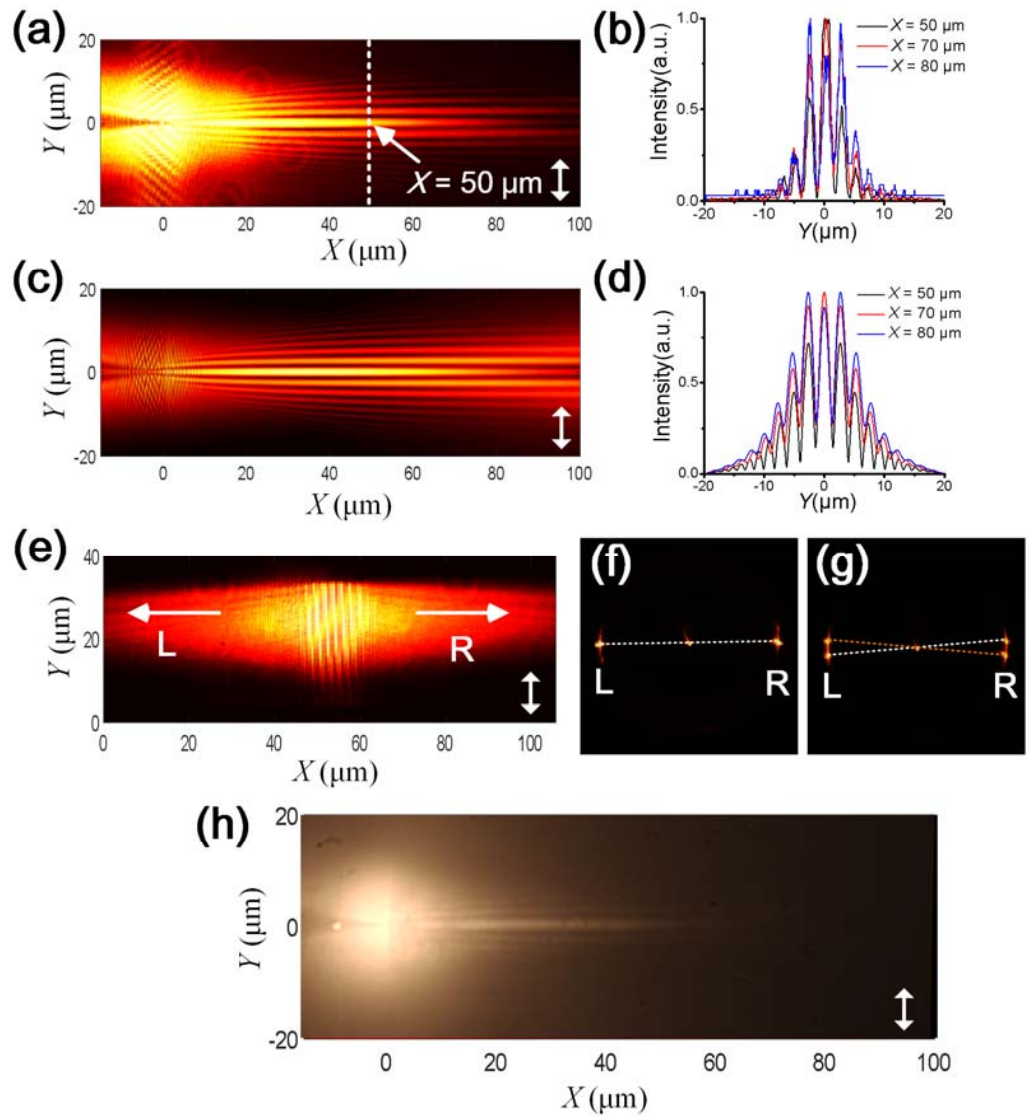

**Figure 2: | DF-BSW at a dielectric/water interface.** (a) Recorded DF-BSW image on the FFP of the LRM. (c) Simulated field distribution of the DF-BSW. (b) Recorded beam shapes at different transverse positions ($X$ = 50, 70, and 80 μm). (d) Simulated beam shapes at the same positions ($X$ = 50, 70, and 80 μm). (e) Recorded FFP-LRM image of a BSW launched by a single grating. The grating's direction (long axis of the grating line) is vertical, so the propagating direction of the launched BSW is perpendicular to the grating's direction. (f) Corresponding BFP-LRM image of the image in panel (e), where the left and right spots represent the BSW propagating to left and right,

respectively. (g) Corresponding BFP-LRM image of the image in panel (a), where the right two spots represent the two BSW plane waves propagating to the right; these then interfere and form a DF-BSW. (h) FFP-LRM fluorescence image of the DF-BSW; the water is doped with the dye Fluorescent red 646 reactive (which can be excited by a laser with a wavelength of 633 nm). A long pass filter (cut-on wavelength at 650 nm) was used to reject the excitation wavelength. The double-headed white arrows in panels (a), (c), (e), and (h) indicate the orientation of the incident polarization.

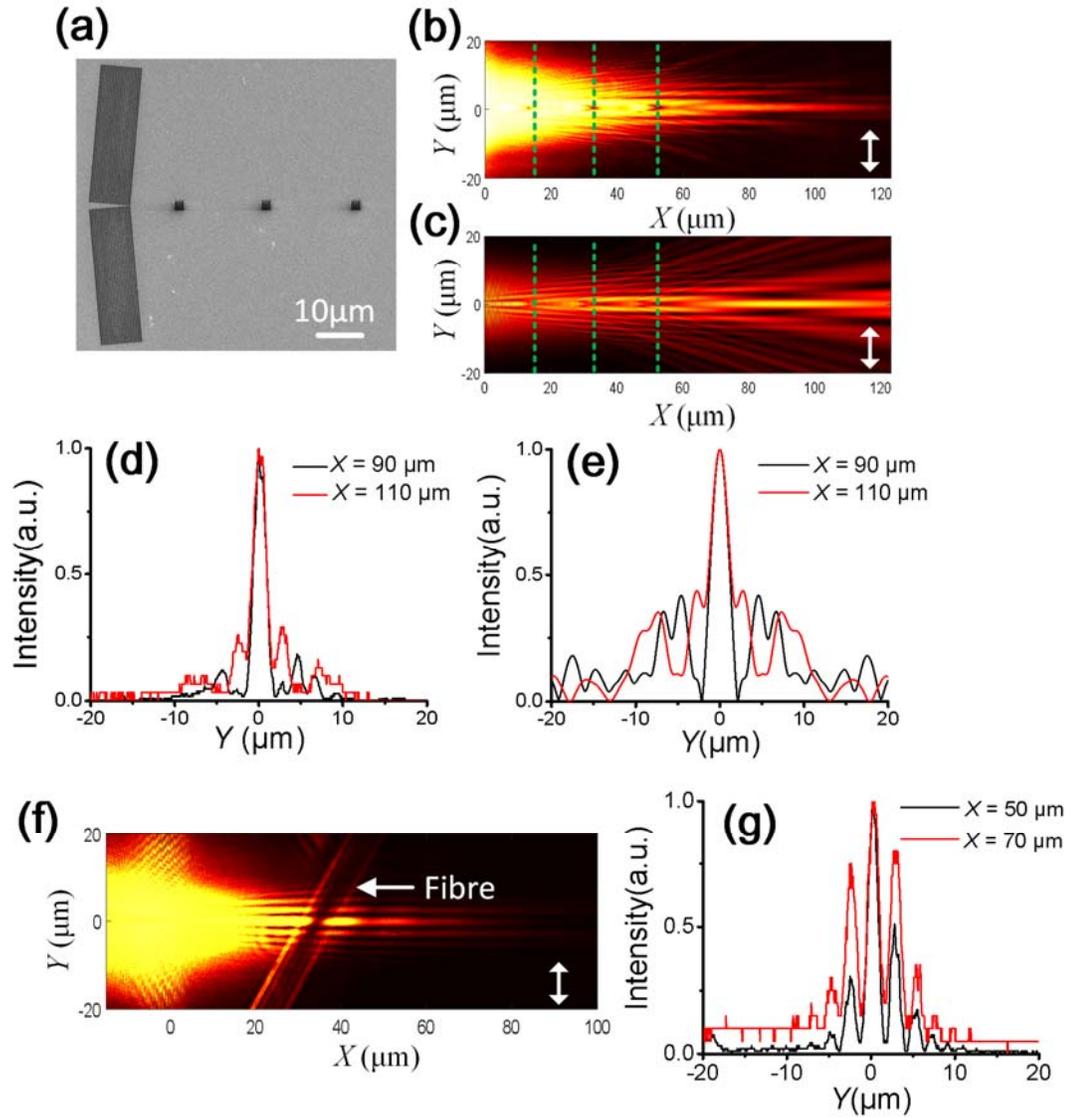

**Figure 3: | Self-healing of the DF-BSW at the dielectric/water interface.** (a) SEM image of the dielectric gratings and three square obstacles inscribed via FIB on the top SiO₂ layer. The distance between the obstacles was 18 µm, and the distance between the last obstacle and the starting point S was 52 µm. The FFP-LRM image (b) and the simulated image (c) of the DF-BSW shows it propagating around the three obstacles. The dashed lines in panels (b) and (c) show the positions of the obstacles. Panel (d) shows the LRM recorded beam shapes at different positions ($X$ = 90 and 110 µm). Panel (e) shows the simulated beam shapes at the same positions ($X$ = 90 and 110 µm). Panel

(f) shows an FFP-LRM image of the BSW propagating across a silica fiber attached to the dielectric multilayer, while panel (g) shows the beam shape of the BSW after the fiber at X = 50 and 70 µm. The double-headed white arrows in panels (b), (c), and (f) indicate the orientation of the incident polarization.

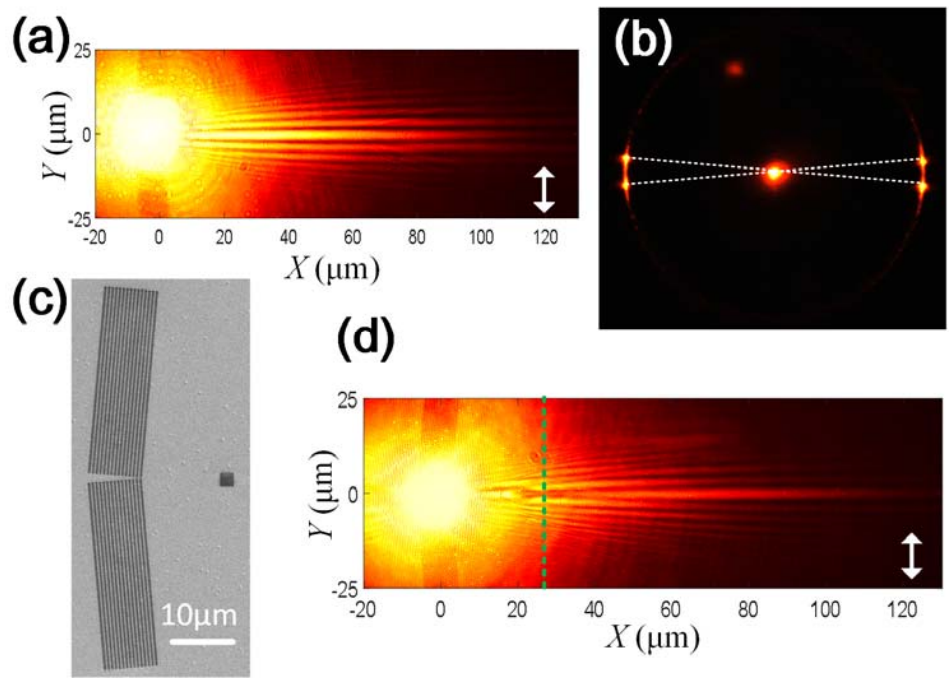

**Figure 4: |DF-BSW at the dielectric/air interface.** (a) FFP-LRM image of the DF-BSW, and (b) corresponding BFP image of (a). (c) SEM image of the gratings with a single obstacle. The cross-angle of the gratings was 170°. The period of the grating was 590 nm. (d) Corresponding FFP-LRM image of (c) in which the dashed line represents the position of the obstacle (X = 28 µm); the size of the square obstacle was 2 µm × 2 µm. The double-headed white arrows in panels (a) and (d) indicate the orientation of the incident polarization.

# Diffraction-Free Bloch Surface Waves

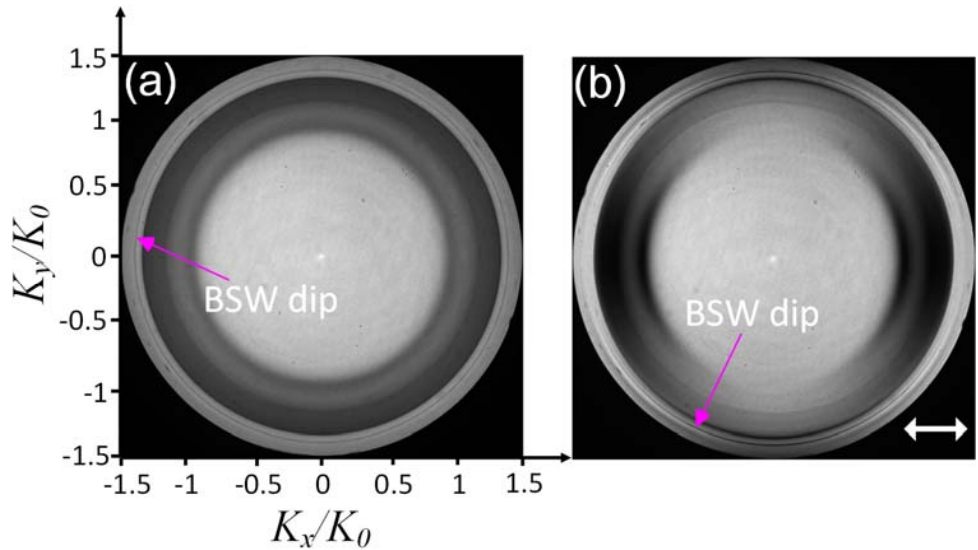

**Supplementary Figure 1 | BFP images of the white light reflected from the dielectric multilayer.**

An expanded white light beam (with a band pass filter that has a center wavelength of 632.8 ± 0.6 nm (FWHM: 3 ± 0.6 nm)) was used to fill the rear aperture of the oil-immersed objective (100× and NA of 1.49). The reflected beam at the BFP of the objective was then imaged onto a CCD camera. The dark dip represents the excitation of the BSW mode, which was equivalent to the dip on the angle-resolved total internal reflected curve measured with a prism setup. From the known N.A of the objective (NA = 1.49), the wavenumber (or effective index) of the BSW mode sustained on the dielectric multilayer could be derived as being about 1.376 $K_0$, so the wavelength of the BSW was about 633 / 1.376 = 460 nm. (a) The reflected BFP image and (b) the reflected BFP image with a polarizer before the CCD. The orientation of the dark arcs on (b) indicates that the BSW on the dielectric multilayer (Fig. 1a) can be excited by TE polarized light. The double-headed white arrow indicates the orientation of the incident polarization.

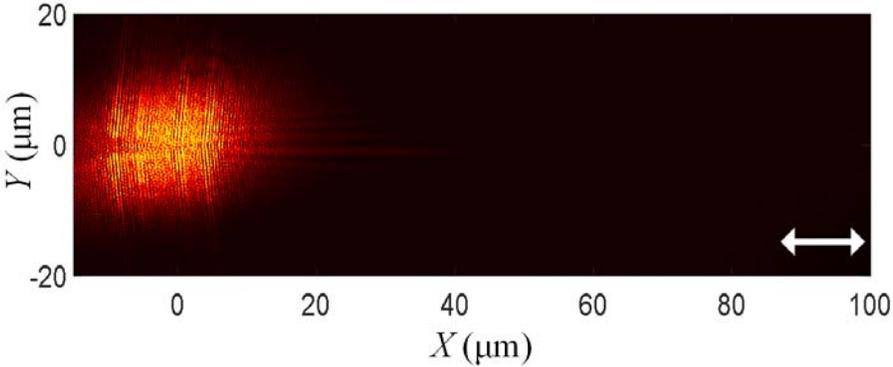

**Supplementary Figure 2 | Incident polarization for the excitation of the DF-BSW.** The FFP-LRM image for a X-polarized (TM polarized) incident laser beam. The double-headed white arrow indicates the orientation of the incident polarization. Under this incident polarization, no DF-BSW could be observed.

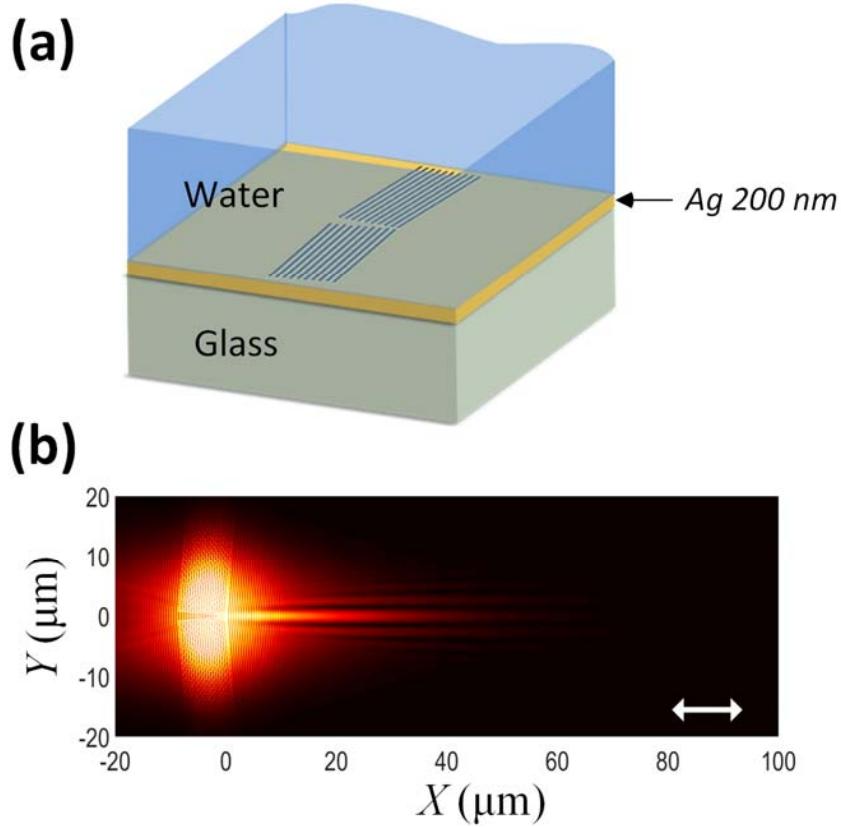

**Supplementary Figure 3 | Numerical simulation of the diffraction-free SPP in water at a 633 nm wavelength.** (a) Simulation model. The refractive index of the Ag at 633 nm was n = 0.135 + 3.988i. The period of the grating was 449 nm and the thickness of the silver film was 200 nm. The sample surface was covered with water. (b) Simulated electric intensity distribution of the launched diffraction-free SPPs wave. The beam waist was 10 μm wide, the same for the DF-BSW. The double-headed white arrow indicates the orientation of the incident polarization. The incident polarization for the ND-SPP was different from that used for the DF-BSWs in Figures 2, 3, and 4.

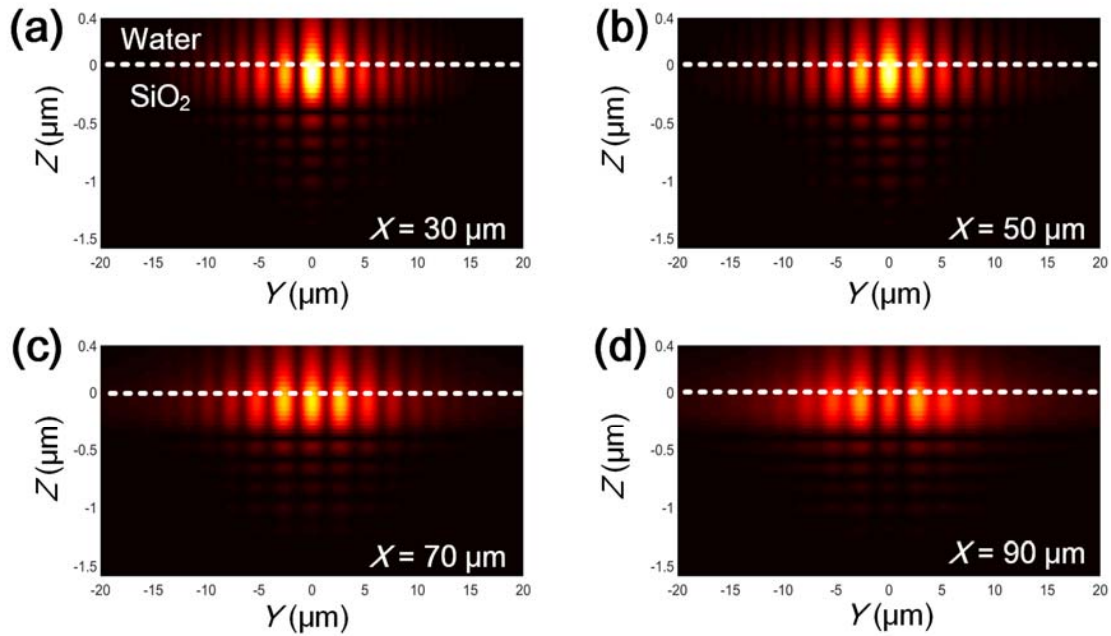

**Supplementary Figure 4 | Simulated electric field distribution of the DF-BSW in the Y–Z planes at different positions of X.** (a) $X$ = 30 µm, (b) $X$= 50 µm, (c) $X$= 70 µm, and (d) $X$= 90 µm. The white dashed line represents the interface between water and the top $SiO_2$ layer. The penetration depth of the DF-BSW remains constant during propagation. The penetration depth of the BSW into water can be increased or decreased by tuning the thickness of the top $SiO_2$ layer.

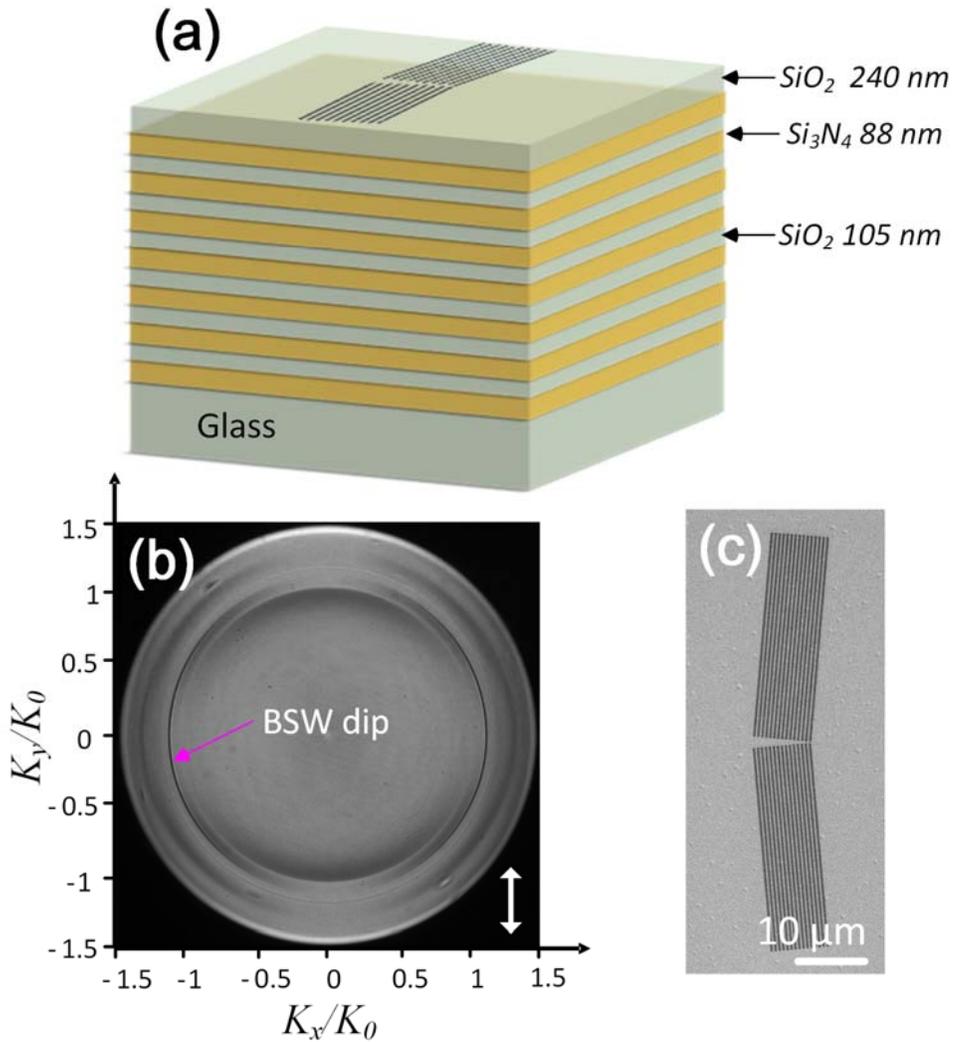

**Supplementary Figure 5 | The multilayer and grating for the DF-BSW on the dielectric/air interface.** (a) Schematic of the dielectric multilayer, the surface was bounded on the top by air. The thickness of the top $SiO_2$ layer was 240 nm, while the other $SiO_2$ layer was 105 nm thick. The thickness of all of the $Si_3N_4$ layers was 88 nm. (b) The reflected BFP images from the multilayer without gratings. The pair of dark arcs (indicated by the BSW dip) represent that BSW mode that can be generated on this multilayer. From the diameter of the arc and the known NA of the objective used in the LRM, the wavelength of the BSW mode can be derived as being 590 nm (the free space incident wavelength is 633 nm). The double-headed white arrow indicates the orientation of the

incident polarization. (c) SEM image of the dielectric gratings fabricated on the top SiO$_2$ layer (the period was 590 nm, which was equal to the BSW wavelength sustained on this multilayer). The cross angle of the two gratings was 170°.